\documentclass{edm_template}
\usepackage{color}
\usepackage{dirtytalk}
\usepackage{multirow}
\usepackage{boldline}
\usepackage{amsmath}
\begin{document}

\title{Predicting Performance on MOOC Assessments using Multi-Regression Models}

%
%
%
%
%

\numberofauthors{3} 
%
\author{
%
%
  \alignauthor
Zhiyun Ren\\
       \affaddr{George Mason University}\\
       \affaddr{4400 University Dr,}\\
       \affaddr{Fairfax, VA 22030}\\
       \email{zen4@masonlive.gmu.edu}
\alignauthor
Huzefa Rangwala\\
       \affaddr{George Mason University}\\
       \affaddr{4400 University Dr,}\\
       \affaddr{Fairfax, VA 22030}\\
       \email{rangwala@cs.gmu.edu}
\alignauthor Aditya Johri\\
       \affaddr{George Mason University}\\
       \affaddr{4400 University Dr,}\\
       \affaddr{Fairfax, VA 22030}\\
       \email{johri@gmu.edu}
}
\date{30 July 1999}

\maketitle

\begin{abstract}

%
%
The past few years has seen the rapid growth 
of data mining approaches for the analysis of 
data obtained from Massive Open Online Courses (MOOCs). 
The objectives of this 
study are to develop approaches to predict the scores a student may achieve
on a given grade-related assessment based on information, considered 
as prior performance or prior activity in the course.
We develop a personalized linear multiple regression (PLMR) model to predict the grade for a student, prior to attempting the assessment activity. The developed model is real-time and tracks the participation of a student within a MOOC (via click-stream server logs) and predicts the performance of a student on the next assessment within the course offering.  We perform a comprehensive set of experiments on data obtained from three openEdX MOOCs via a Stanford University initiative. Our experimental results show the promise of the proposed approach in comparison to baseline approaches and also helps in identification of key features that are associated with the study habits and learning behaviors of students.
\end{abstract}

%

\keywords{Personalized Linear Multi-Regression Models, MOOC, Performance prediction} 

\section{Introduction}

Since their inception, Massive Open Online 
Courses (MOOCs) have aimed at delivering online learning on a wide variety of 
topics to a 
large number of participants across the world. Due to the low cost (most times zero) 
and lack of entry barriers (e.g., prerequisites or 
skill requirements)  for the participants, large number 
of students enroll 
in MOOCs but only a small fraction  of them keep 
themselves engaged in the learning materials 
and participate in the various activities associated 
with the course offering 
such as viewing the video lectures, studying the material, 
completing the various quizzes and homework-based 
assessments.

Given, this high attrition 
rate and potential of MOOCs to deliver low-cost 
but high quality education,  several 
researchers have analyzed the 
server logs associated with these MOOCs
to determine the factors associated with  
students  dropping out.  Several predictive methods
have been developed to predict when a 
participant will drop out 
from a MOOC ~\cite{greene2015predictors,hughes2015utilization, jiang2014predicting, whitehill2015beyond, yang2013turn, ye2014early,ye2015behavior}.
Using self reported surveys, 
studies have determined
the different motivations for students enrolling and 
participating in a MOOC. Participants 
enroll in a MOOC sometimes to learn
a subset of topics within the curriculum, sometimes 
to earn degree certificates for future career 
promotion or college credit, social experience or/and
exploration of 
free online education \cite{onah2015learners}. 
Students with similar motivation 
have different learning outcomes from a MOOC based on the number of 
invested hours, prior 
education background, knowledge and skills \cite{greene2015predictors}.

In this paper, we present models to 
predict a student's future performance for 
a certain assessment activity  witin a MOOC. 
Specifically, we develop an approach
based on personalized linear multi-regression (PLMR)  
to predict the performance of a student as they attempt 
various graded activities (assessments) within the MOOC. 
This approach was previously studied within the context of predicting 
a student's performance based on graded activities within a traditional university
course with data extracted from a learning management system (Moodle) \cite{elbadrawy2014personalized}.
The developed 
model is real-time and tracks the participation of a 
student within a MOOC (via click-stream server logs) 
and 
predicts the performance of a student on the next assessment
within the course offering. Our approach also 
allows us to capture the varying 
 studying  patterns associated with different students, and responsible for 
their performance. 
We evaluate our predictive model on 
three MOOCs offered using the OpenEdX platform and made available for learning 
analytics research via  the 
Center for Advanced Research through
Online Learning at Stanford University \footnote{datastage.stanford.edu}.

We extract features that seek to identify the learning behavior 
and study habits for 
different students.  These features capture the 
various interactions that show engagement, effort, learning and behavior 
for a given student participating in studying; by viewing the various 
  video and text-based materials available within the 
  MOOC offering coupled 
  with student attempts on graded and 
  non-graded activities like quizzes 
  and homeworks. 
%
%
%
Our experimental evaluation shows accurate 
grade prediction for different types of homework assessments in comparison to 
baseline models. 
Our approach also identifies the features found to be useful 
for predicting an accurate homework grade.

 Baker et. al  \cite{baker2008improving} have presented systems that can adapt
  based on predictions of 
  future student performance, 
  and they were
   able to incorporate  interventions, which were effective in improving student
    experiences within 
    Intelligent Tutoring Systems (ITS).
Inspired by this prior work, tracking 
student performance within  a MOOC, allows  personalized feedback for
high performing and low performing students; motivating students to 
stay on track and achieve their educational goals. It  also  provides
feedback to the MOOC instructor about the usage of different 
course materials and helps in improving the MOOC offering. 
%

%


\section{RELATED WORK}

Several researchers have focused on the 
analysis of education data (including MOOCs), in
an effort to 
understand the characteristics of 
student learning behaviors and motivation within this education model \cite{pena2014educational}. Boyer et. al. \cite{boyer2015transfer} focus on 
the stopout prediction problem within MOOCs; by 
designing a set of processes using information from 
previous courses and the previous weeks of the 
current course. 
Brinton et. al. \cite{brinton2015mooc} developed an approach to 
predict if a student answers a question correct on the 
first attempt via click-stream information and social learning
networks. Kennedy et. al. \cite{kennedy2015predicting} 
analyzed the relationship between a student's 
prior knowledge on end-of-MOOC performance. 
Sunar et. al. \cite{sunar2015analysing} developed an approach 
to  predict the possible interactions between peers participating in a MOOC. 

Most similar to our proposed work, 
Bayesian Knowledge Tracing (BKT) \cite{pardos2011kt} has been adapted to 
predict whether a student can get a MOOC assessment correct or not.
BKT was first developed \cite{corbett1994knowledge} for modeling the 
evolving knowledge states of students monitored within Intelligent 
Tutoring Systems (ITS).  Pardos et. al. 
proposed a model \say{Item Difficulty Effect Model} (IDEM)
that incorporates  the difficulty levels 
of different questions and modifies the original BKT by 
adding an \say{Item} node to every question node.  By identifying 
the  challenges associated with 
modeling MOOC data, the IDEM approach and extensions  that involve 
splitting questions into several sub-parts and incorporating resource (knowledge) information \cite{pardos2013adapting}
are considered state-of-the-art MOOC assessment 
prediction 
approaches and referred as KT-IDEM. 
However, this approach can only predict a binary value grade.  In contrast, the model proposed in this paper is 
able to predict both, a continuous and a binary grade. 

Within learning analytics literature, outside of MOOC analysis,
predicting student performance is a popular and extensive topic. 
Wang et. al. \cite{wang2015smartgpa} performed a study to predict student's performance by capturing 
data relevant to study habits and learning behaviors
from their smartphones. Specific examples of data captured include 
location, time, ambient noise and social activity.  Coupled with
self-reported information, this work captured the influence of a student's
daily activity on the academic performance. 
Elbadrawy et. al. \cite{elbadrawy2014personalized} proposed the use of 
personalized linear multi-regression models to predict student
performance in a traditional university by extracting data from 
course management systems (Moodle). With a particular 
membership vector for each student, the model was able to capture
personal learning behaviors
and outperformed several baseline approaches. Our 
 study  focuses on  MOOCs, which presents
 different assumptions, challenges and features in comparison to a  traditional university 
 environment.

\section{METHODS}
\subsection{Personal Linear Multi-Regression Models}
We train a personalized linear multi-regression (PLMR) model \cite{elbadrawy2014personalized} to predict student
performance within a MOOC. Specifically,  the 
grade $\hat{g}_{s,a}$ for a student $s$ in an assessment activity $a$ is predicted as follows:
\begin{equation}\label{eq1}
\begin{aligned}
\begin{split}
\hat{g}_{s,a}  & = b_s+p_{s}^{t}Wf_{sa}\\
 &  = b_s+\sum_{d=1}^{l}(p_{s,d}\sum_{k=1}^{n_F}f_{sa,k}w_{d,k}),
\end{split}
\end{aligned}
\end{equation}
where $b_s$ is bias term for student $s$, $f_{sa}$ is the feature vector of an interaction between student $s$ and activity $a$. The features extracted from the MOOC server logs are described in the next Section. $n_F$ is
the length of $f_{sa}$, indicating the dimension of our feature space. $l$ is the number of 
linear regression models, $W$ is the coefficient matrix 
of dimensions $l \times n_F$ that holds the coefficients of the $l$ linear regression models, and $p_s$ is a vector of length $l$ that 
holds the memberships of student $s$ within the $l$ different regression models \cite{elbadrawy2014personalized}. Using lasso \cite{tibshirani1996regression}, we solve
the  following optimization problem:

\begin{equation}\label{eq2}
\underset{(W,P,B)}{minimize}\ L(W,P,B)+\gamma  (\left \| P \right \|_F+\left \| W \right \|_F),
\end{equation}

where $W$, $P$ and $B$ denote the feature weights, student memberships and bias terms, respectively. The loss
function $L(\cdot)$ is the least square loss  for regression problems.
$\gamma  (\left \| P \right \|_F+\left \| W \right \|_F)$ is a regularizer that controls the 
model complexity by controling the values of feature weights and student memberships. 
Tuning the scalar $\gamma$  prevents model from over-fitting.


\subsection{Feature Description}

We extract features from MOOC server logs and 
formulate the PLMR model 
to predict real-time assessment grade for a given student.
%
Figure \ref{fig:figure3} shows the 
various
activities, generally available  within a MOOC. Fig \ref{fig:figure3}  (a) shows 
that each homework has corresponding quizzes, 
each of which has its corresponding video as resources for learning. Fig \ref{fig:figure3} (b) shows that while 
watching a video, a student can  have a series of actions. Fig \ref{fig:figure3} (c) shows 
that while studying using a  MOOC, a student can   have several login sessions, 
each of which may involve  watching videos, attempting quizzes and  homework related activities. In order
to capture the latent information behind the click-stream for each student, we extract six types of 
features: (i) session features, (ii) quiz related features, (iii) video related features, (iv) homework related features, (v) time related features and (vi) 
interval-based features. These features constitute the 
feature vector $f_{sa}$ for a student and a homework assessment. 
The description of these features are as follows:

\begin{figure}[th]
\centering
  \includegraphics[width=0.9\columnwidth]{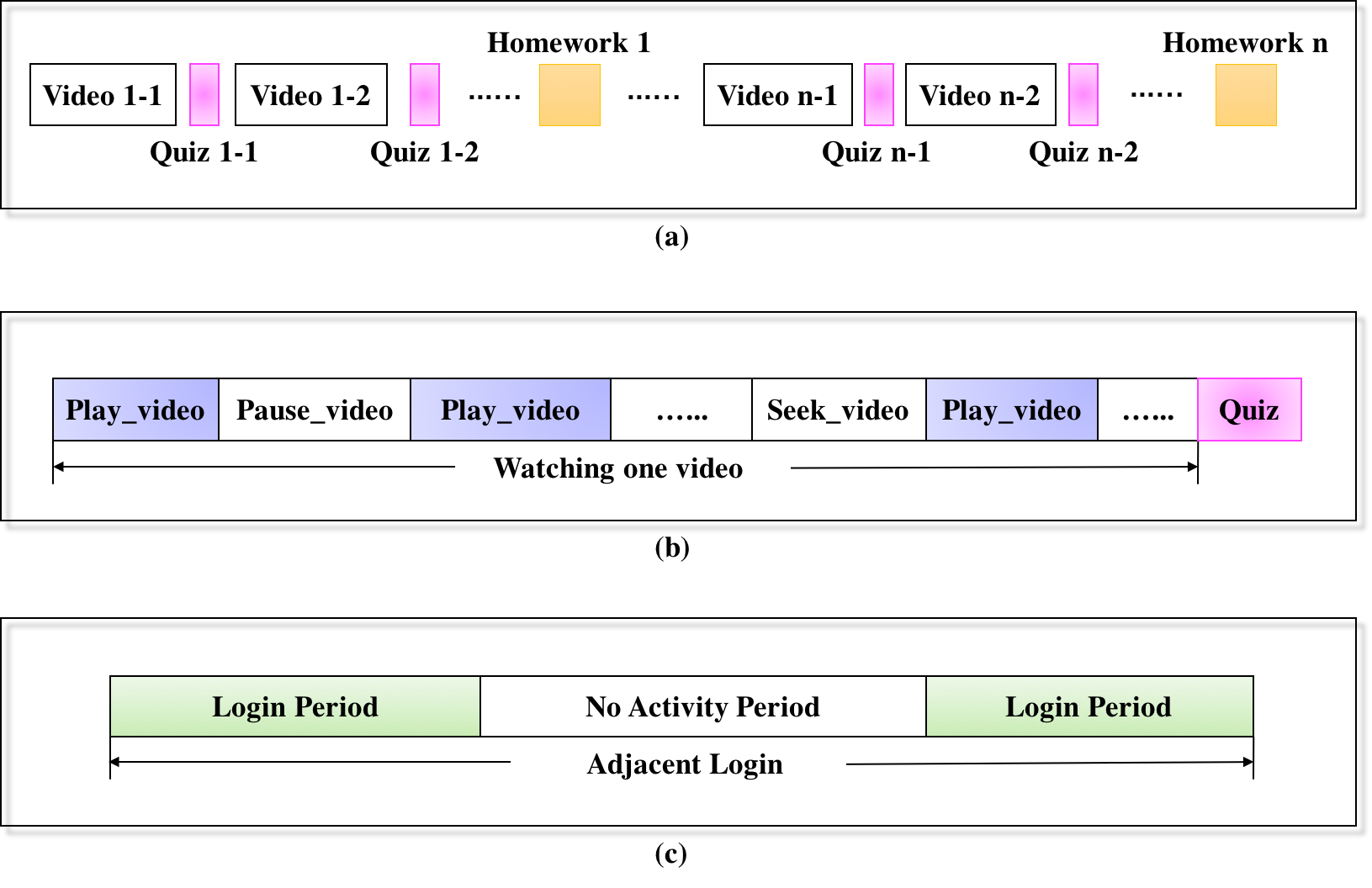}
  \caption{Different activities within a MOOC.}~\label{fig:figure3}
\vspace{-0.8cm}
\end{figure}

\paragraph*{(i) Session features:}
~\\
A single study session is defined by a student login combined with the 
various available study 
interactions that a student may partake in.  Since, students 
do not always log out of a session, we assume that a \say{no activity} period 
  of more 
than one hour constitutes a student logging out of a session. We show a \say{no activity} period for 
a student between tarwo consecutive sessions 
in Fig \ref{fig:figure3} (c).

\begin{itemize} \item \textbf{NumSession} are the the average 
number of daily study sessions 
a student engages in, before a homework attempt. 
\item \textbf{AvgSessionLen} is the average length of each session in minutes. We 
calculate the average study time of a study session by
\begin{equation}\label{eq3}
AvgSessionLen=\frac{Total\ study\ time}{NumSession}.
\end{equation}

\item \textbf{AvgNumLogin} is the percentage of days before a homework attempt that a student logs in to MOOC (or has a session). Students are free to choose when to login and study in a MOOC environment. 
We consider a day as a 
\say{work day} if a student logs into the system and has some study related 
activities; and a  day as \say{rest day} if a student does not 
login and has no study-related 
activities. The rate of \say{work} and \say{rest}  can capture
a student's learning habits and engagement 
characteristics. 
\begin{equation}\label{eq4}
\begin{aligned}
\begin{split}
Avg & NumLogin=  \\
& \frac{\#\ of ``work\ day"}{\#\ of\  ``work\ day"\ + \#\ of\ ``rest\ day"}.
\end{split}
 \end{aligned}
\end{equation}

\end{itemize}
\paragraph*{(ii) Quiz Related features:}
\begin{itemize}
\item \textbf{NumQuiz} are the number of quizzes a student takes before a homework attempt. In the analyzed MOOCs, 
every homework has its corresponding quizzes, and each quiz has its
own corresponding video(s) as shown in Fig \ref{fig:figure3} (a). Students are expected 
to watch the videos and attempt
the quizzes before they attempt
each homework. The number of quizzes a student attempts   reflects the 
student's dedication towards the course material and a factor 
towards performance in a homework.
\item \textbf{AvgQuiz} is the  average number of attempts for each quiz. The MOOCs studied in this paper  allow 
unlimited attempts on a quiz.

\end{itemize}
\paragraph*{(iii) Video Related features:}
\begin{itemize}
\item \textbf{VideoNum} denotes  the number of distinct video sessions for  a student before a homework attempt.
\item \textbf{VideoNumPause} is the average number  of pause actions per video. There are several actions associated with viewing  videos, 
including ``pause video", ``play video", ``seek video" and  ``load video". 
Tracking these student actions allows for 
capturing a student's focus level and learning habits. 
If a student pauses a video several times, we assume that the student 
is thinking about the content and stops to research other materials. However,
we can also assume that the student may pause several times due to a  lack of focus.   On the contrary, if a student does not 
pause a video during the watching time, it could suggest   
that either the student understands everything or  is distracted and 
loses focus.

\item \textbf{VideoViewTime} is the total video viewing time. Different videos have different lengths. Students can also stop watching the video in the middle. We calculate the whole video watching time instead of average watching time for each video.
\item \textbf{VideoPctWatch}  In a large amount of cases, students do not 
watch the complete video session. As such, we calculate the average fraction of watched part out of the total  video length.
\end{itemize}
\paragraph*{(iv) Homework Related features:}
\begin{itemize}
\item \textbf{HWProblemSave} is the average number of saves (event coding is \say{problem$\_$save}) 
for each homework assessment.  Students only have one chance to do the homework and the 
  action \say{problem$\_$save} is for the situation that the students 
  have already done some part of a homework or all of it, but are 
  not ready to submit it for assessment and grading. Students 
  may save the homework and submit it after a few days during 
  which time they may
  check the homework several times. As such,  the \say{problem$\_$save} event 
  reflects studying patterns  for students.
\end{itemize}
\paragraph*{(v) Time Related features:}
\begin{itemize}
\item \textbf{TimeHwQuiz} is the time difference 
between a homework attempt and the last quiz a student attempts before that homework. Quizzes
help student understand the material. The 
corresponding quizzes of a homework might 
have similar questions as with the homework. Attempting a quiz helps 
students recall the knowledge and may lead to improved 
performance in the upcoming homework assessment.
\item \textbf{TimeHwVideo} is the time difference between a homework attempt 
and the last video a student watches before that homework.
\item \textbf{TimePlayVideo} is the average fraction 
of study sessions that have "play video" over all the 
study sessions. 
%
We calculate TimePlayVideo by:
\begin{equation}\label{eq5}
\frac{\#\ of\ study\ sessions\ that\ have\ ``play\ video"}{\#\ of\ all\ study\ sessions}.
\end{equation}
\item \textbf{HwSessions} is the number of sessions that have 
homework related activities (save and submit). Although students 
have only one chance to submit a 
homework, they have sufficient time to review saved homework's answers. As such, 
saving  and submitting the same homework could occur in different 
sessions and possibly   different days.
\end{itemize}

\paragraph*{(vi) Interval-Based features:}
~\\
Several of the features described above are cumulative in nature and aggregated from the time (session) the student 
signs on to participate in a MOOC. However, we also want to capture the features aggregated between consecutive homeworks. 
It is expected that there will be some changes in student learning related activities once, they know the former homework's grade. 
For 
example, some students will study harder if they do not perform well 
on a previous homework. So we extract a 
group of features that represents activities between two 
consecutive homeworks.

\begin{itemize}
\item	 \textbf{IntervalNumQuiz}: denotes the number of quizzes the student takes between two homeworks.
\item	 \textbf{IntervalQuizAttempt}: is the average number of quiz attempts  between two homeworks.
\item \textbf{IntervalVideo}: is the number of videos a student watches between two homeworks.
\item \textbf{IntervalDailySession}: is the average number of sessions per day between two homeworks.
\item \textbf{IntervalLogin}: is the percentage of login days between two homeworks.
\end{itemize}

We also use the cumulative grade (so-far) on quizzes and homeworks for a student as a feature and denote it by \\ \textbf{Meanscore}. For 
our baseline approach we only consider the averages computed on the 
previous homeworks. 
\begin{figure}[th]
\centering
  \includegraphics[width=0.7\columnwidth]{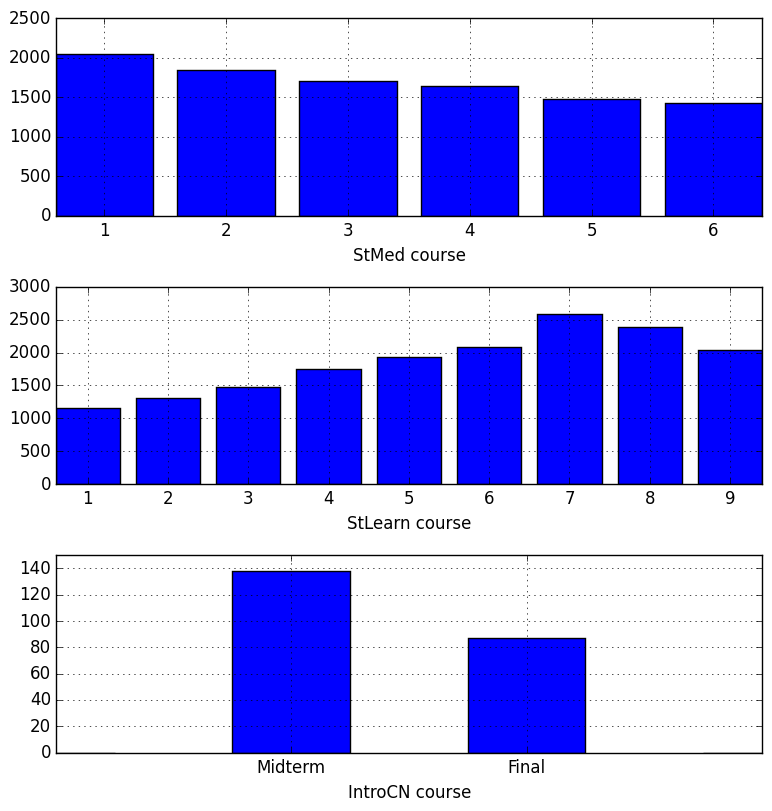}
  \caption{Distribution of students attempting each Assessment. \tiny{StMed, StLearn and IntroCN had 6, 9 and 2 assessments, respectively.}}~\label{fig:figure2}
\vspace{-1.5em}
\end{figure}

\section{EXPERIMENTS}
\subsection{Datasets}
We evaluated our methods on three MOOCs: \say{Statistics in Medicine} (represented as StMed in this paper) taught in Summer 2014, \say{Statistical Learning} (represented as StLearn in this paper) taught in Winter 2015 and \say{Introduction to Computer Networking} (represented as IntroCN in this paper) taught in Spring 2015.

\textbf{StMed:} This dataset includes server logs tracking 
information about a student viewing  video lectures,  checking text/web articles,  attempting quizzes and homeworks (which are graded). Specifically, this MOOC   contains 9 learning units with 111 assessments, including 79 quizzes, 6 homeworks and 26 single questions. The course had 13,130 students enrolled, among which 4337 students submitted at least one assignment (quiz or homework) and had corresponding scores, 1262 students have completed part of the six homeworks and 1099 students have attempted 
all the homeworks. 193 students attempted all the 79 quizzes and six homeworks. This course had 131 videos and 6481 students had 
video related activity.\\

  \begin{figure*}[th]
\centering
  \includegraphics[width=0.85\textwidth]{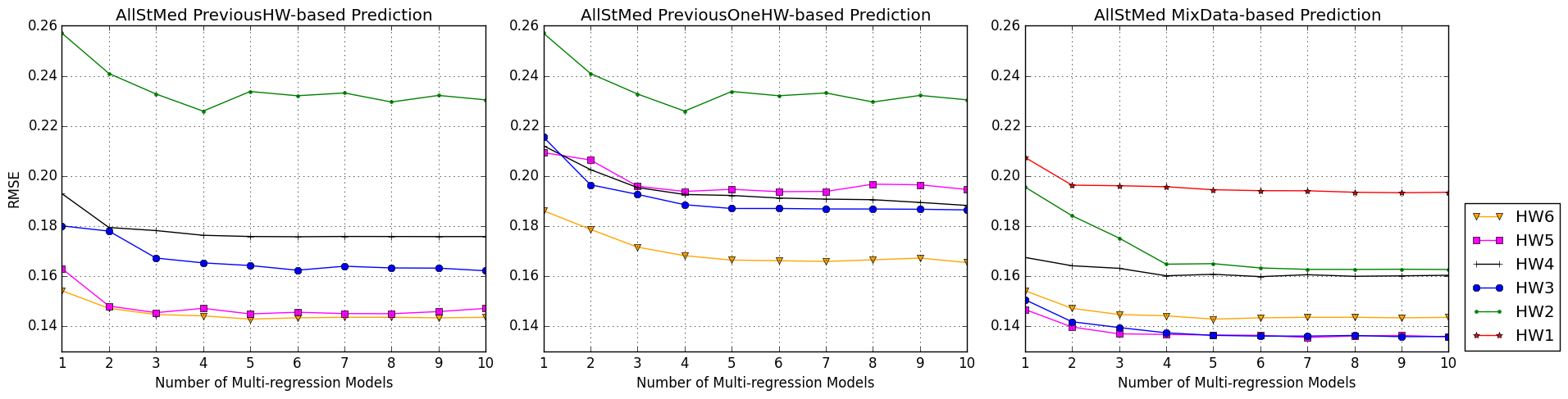}
  \caption{AllStMed Prediction Results. RMSE ($\downarrow$ is better). }~\label{fig:figure4}
\end{figure*}
\vspace{-1em}
\begin{figure*}[th]
\centering
  \includegraphics[width=0.85\textwidth]{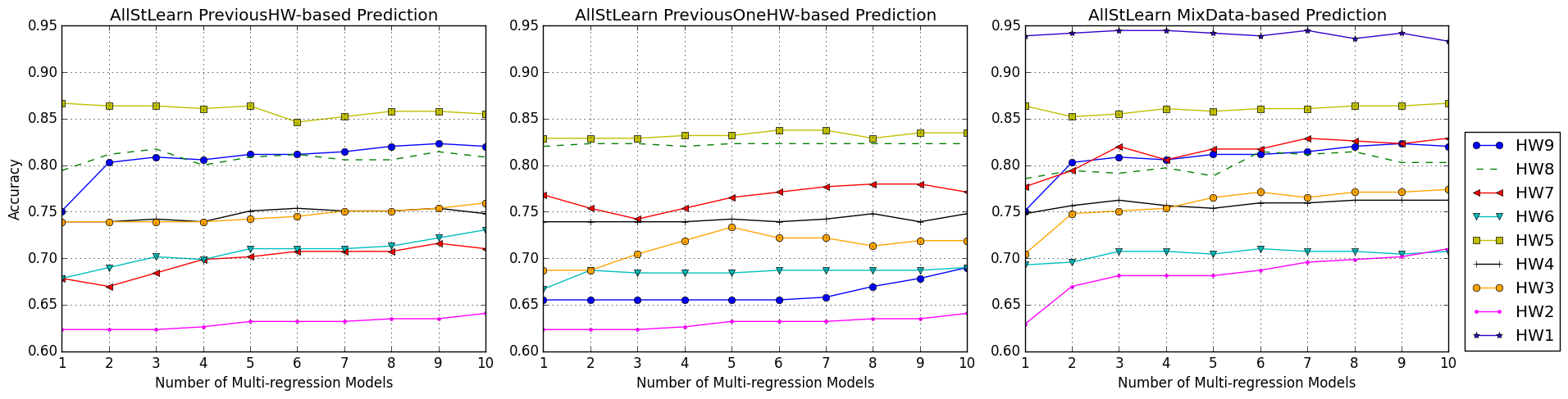}
  \caption{AllStLearn Prediction Results. Accuracy ($\uparrow$ is better).}~\label{fig:figure5}
\end{figure*}

\textbf{StLearn:} This course had ten units. Except the first one, all units have quizzes and end of unit homeworks, which add up to 103 assessments in total. 52,821 students enrolled in this course, and 4987 students had assessment activities, 3509 students  attempted a subsets of the available 
homeworks while 346 students attempted  all the 9 homeworks, and 118 students attempted 
all the 103 assessments.  The key difference between 
the homeworks in the StLearn in comparison to the StMed  is that homeworks have only one question 
which a student can either get correct or incorrect. As such, scoring in this MOOC is binary instead of continuous. To 
predict whether a student answers a  question correctly, 
we reformulate the regression problem as a classification problem using  a logistic loss function.

\textbf{IntroCN:} This class had 8 units, a midterm exam and a final exam, including 244 assessment activities. 16,395 students enrolled in
this course out of which  3263 students had 
assessment activities, among whom 84 students finished both the midterm and final exams. For this dataset, we predict the a student's 
performance for the final exam with the model trained by the information prior to their midterm exam.\\
 
 Figure \ref{fig:figure2} shows the distribution
 of students attempting the different assessments available 
 across the three MOOCs studied here.

\subsection{Experimental Protocol}

In order to gain a deep insight of 
students' performance in a MOOC, we perform three 
types of experiments.
Given $n$, homework assessments represented as $\{H_1, \ldots, H_n\}$ our 
objective is to predict the score a student achieves in each of the 
$n$ homeworks. 
Depicting the most realistic setting, for the $i$-th homework, $H_i$ we 
define the training set as all 
homework and student pairs  who attempt and have a score for all homeworks up to the 
$H_{i-1}$.  For predicting the score for $H_i$ for a given student, we use all the 
features extracted just before attempting the target homework $H_i$. We refer to 
this as \textbf{PreviousHW-based Prediction}. 
Secondly, for the predicting $i$-th homework $H_i$'s score, 
we use training data of student-homework 
pairs restricted from only the previous one homework i.e.,  $H_{i-1}$. This experiment 
is referred by \textbf{PreviousOneHW-based Prediction}.
Note, in these cases we cannot make any prediction for the first homework ($H_1$) since, 
we do not have any training information for a given student. 
We also formulate an experiment that ignores the sequence of homeworks and makes a prediction 
for the target $H_i$ using training data of student-homework pairs from all  the homeworks except 
  $H_i$ i.e., $\{H_1 \ldots H_n\} - H_i$. This allows for assessment of the models using the most 
  training data available from the MOOC, and does not assume that students should follow the 
  sequence of homeworks as suggested by the instructor.  We refer to this experiment by 
  \textbf{MixData-based Prediction}.

\subsection{Data Partition}
We partition the students for StLearn and StMed
into two groups: the group of 
students who attempt \emph{all} 
the requested homeworks, and the group of students
who finish \emph{few} of the homeworks.
This allows us to 
consider the different motivations and 
expectations of students enrolling in a MOOC. 
 For example, the students who aim to learn in 
a MOOC may watch videos for a long time and not attempt the 
homeworks.  While,
the students who want to achieve a degree 
certificate may not pay so much effort 
in watching the videos but focus on the homework scores. 
We refer to  the 
first group by \say{Partial homeworks accomplished group}, and the 
second group by \say{All homeworks accomplished group}.  We evaluate our models on
the two groups for the \textbf{AllStMed} and \textbf{AllStLearn} datasets. Specifically, we name the 
four group of students as \textbf{AllStMed}, \textbf{AllStLearn}, \textbf{PartialStMed} and 
\textbf{PartialStLearn} based on their group and MOOC class. 

For the \textbf{IntroCN} course,
both midterm exams and final exams have a 
certain amount of quizzes available for practice 
beforehand. For this dataset our goal is to predict the 
final exam prediction score. As such, we include 
all students who attempt this final exam in our analysis.




%
%
\begin{table}[th]
\begin{center}
\begin{tabular}{ |c|c|c| } 
\hline
HW\#  & PLMR & Meanscore\\
\hline
2 & \textbf{0.230} & 0.248\\
\hline
3 & \textbf{0.162} & 0.176\\
\hline
4 & \textbf{0.176} & 0.196\\
\hline
5 & \textbf{0.144} & 0.156\\
\hline
6 & \textbf{0.143} & 0.150\\
\hline
Avg & \textbf{0.171} & 0.185\\
\hline
\end{tabular}
\end{center}
  \caption{PreviousHW-based RMSE Performance (RMSE) comparison for AllStMed.}~\label{tab:table1} \end{table}

\begin{table*}[th]

\begin{center}
\begin{tabular}{ |c|c|c|c|c|c|c| } 
\hline
\multirow{3}{*}{HW\#} &
   \multicolumn{3}{c|}{Accuracy ($\uparrow$)} &
    \multicolumn{3}{c|}{$F_1$ ($\uparrow$)} \\
   \cline{2-7}
  & \multirow{2}{*}{PLMR} & \multicolumn{2}{c|}{Baseline} & 
      \multirow{2}{*}{PLMR} & \multicolumn{2}{c|}{Baseline}  \\
      \cline{3-4}
      \cline{6-7}
   & & Meanscore & KT-IDEM & & Meanscore & KT-IDEM\\
\hlineB{2}
 2 & 0.641 & \textbf{0.646} & 0.623 & 0.775 & \textbf{0.777} & 0.768 \\
\hline
3 & \textbf{0.760} & 0.580 & 0.681 & \textbf{0.821} & 0.805 & 0.810\\
\hline
4 & \textbf{0.754} & 0.710 & 0.739 & 0.838 & 0.706 & \textbf{0.850}\\
\hline
5 & \textbf{0.867} & 0.809 & 0.829 & \textbf{0.920} & 0.880 & 0.906\\
\hline
6 & \textbf{0.730} & 0.678 & 0.667 & \textbf{0.808} & 0.776 & 0.800\\
\hline
7 & 0.716 & 0.675 & \textbf{0.730} & \textbf{0.887} & 0.878 & 0.844\\
\hline
8 & \textbf{0.817} & 0.762 & 0.817 & \textbf{0.903} & 0.849 & 0.886\\
\hline
9 & \textbf{0.823} & 0.794 & 0.777 & \textbf{0.864} & 0.856 & 0.853\\
\hline
Avg & \textbf{0.764} & 0.707 & 0.759 & \textbf{0.852} & 0.816 & 0.848\\
\hline
\end{tabular}
\end{center}
  \caption{PreviousHW-based prediction performance comparison for AllStLearn group.}~\label{tab:table3}
\end{table*}

%

%
%
\begin{table}[h]

\begin{center}
\begin{tabular}{ |c|c|c|c| } 
\hline
HW\# & PLMR & KT-IDEM\\
\hline
2 & \textbf{0.641} & 0.623 \\
\hline
3 & \textbf{0.733} & 0.681 \\
\hline
4 & \textbf{0.748}  & 0.739 \\
\hline
5 & \textbf{0.838}  & 0.829 \\
\hline
6 & \textbf{0.690} & 0.667 \\
\hline
7 & \textbf{0.780} & 0.730\\
\hline
8 & \textbf{0.823} & 0.823\\
\hline
9 & \textbf{0.690} & 0.655 \\
\hline
Avg & \textbf{0.743} & 0.718 \\
\hline
\end{tabular}
\end{center}
  \caption{PreviousOneHW-based Prediction Performance (Accuracy score) comparison for AllStLearn}~\label{tab:table5}
\end{table}
%
%


\begin{figure}[h]
\centering
  \includegraphics[width=0.6\columnwidth]{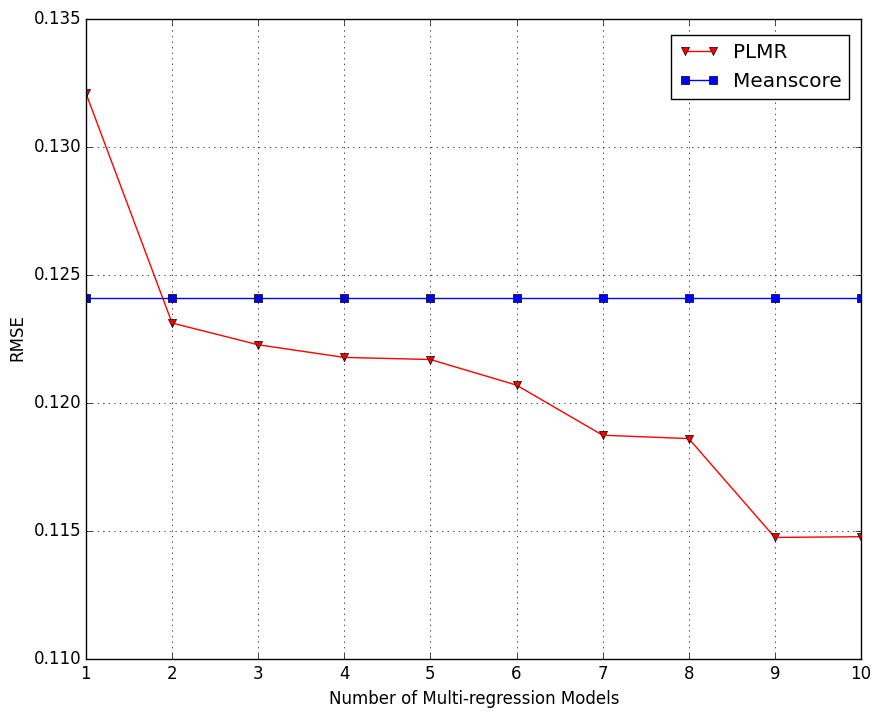}
  \caption{IntroCN Prediction Results. \tiny{RMSE ($\downarrow$ is better).}  ~\label{fig:figure10}}
\vspace{-1em}
\end{figure}

\begin{figure}[th]
\centering
  \includegraphics[width=0.8\columnwidth]{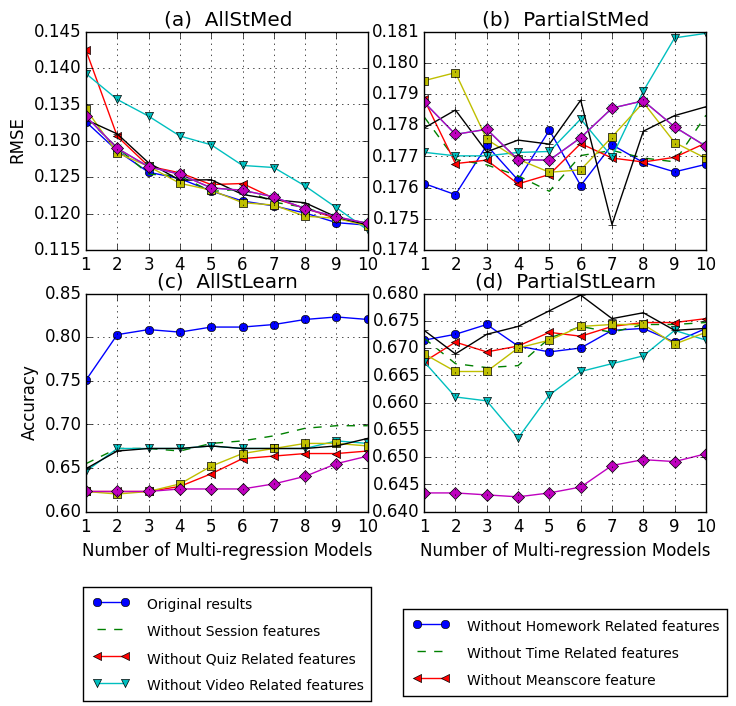}
  \caption{Predictive Performance with Removal of Feature Types. }~\label{fig:figure7}
\vspace{-1em}
\end{figure}

\subsection{Evaluation Metrics}
StMed and IntroCN courses have
continuous scores for a homework or 
an exam, which are scaled between 0 and 1. However, 
the  homework score is binary in the
StLearn course, indicating 
whether the student answers a  question correctly or incorrectly.  For StLearn, we use a logistic loss 
and formulate a classification problem instead of the regression problem 
as done for the StMed and IntroCN courses. 
To evaluate the performance 
of our approach, we use the root mean squared error (RMSE) as the metric of choice for regression problem. For 
classification problem, we use accuracy and the F1-score (harmonic 
mean of precision and recall), known to be a suitable metric for imbalanced datasets.

\subsection{Comparative Approaches.}
In this work, we compare the performance of our proposed methods with 
two different competitive baseline approaches.
\paragraph*{1. \textbf{Average grade of the previous homeworks}} We calculate the mean score of a given student's previous 
homeworks to  predict their future performance and is denoted as Meanscore. We use this method to compare our prediction results on StMed.
\paragraph*{2. \textbf{KT-IDEM \cite{pardos2011kt}}} KT-IDEM is a modified version of original BKT model. By adding an \say{item} node to every question node, the model assigns different probability of \say{slip} and \say{guess} to different questions, due to uneven difficulty each question has. Since this model can only predict a 
binary value grade, we use this model to compare our prediction results on StLearn.

\section{Results and Discussion}
\subsection{Assessment Prediction Results}
Figures \ref{fig:figure4}  and \ref{fig:figure5}
show the prediction results with varying 
number of regression models for the 
AllStMed and AllStLearn  MOOCs, respectively. 
Analyzing Figure \ref{fig:figure4} we observe that as the number 
of regression models increase the RMSE metric goes lower and use 
of five models seems to be good choice for all the different homeworks. 
Comparing the PreviousHW- and PreviousOneHW-based results, we notice that predictions 
for all the homeworks (HW3, HW4, HW5, and HW6) 
  benefits from using all the available training data prior to those homeworks i.e., to 
  predict grade for $H_{i}$ it is better to use training information extracted from $H_{1} \ldots H_{i-1}$ rather 
  than just $H_{i-1}$.
 Comparing the MixData-based prediction results we notice the improved performance for all the homeworks in 
 comparison to the PreviousHW-based prediction results. 
%
Similar observations can be made while analyzing the prediction results for the AllStLearn cohort which includes
 nine homework correct/incorrect binary assessments. Figure \ref{fig:figure5} shows the accuracy scores (higher is better) 
 for the three experiments.  For the PreviousOneHW- and PreviousHW-based experiments HW5 shows the 
   best prediction results. This suggests that in the middle of a MOOC, students tend to have stable study activities and the performance is more predictable than other phases. Other interesting 
   observations include, that for the MixData-based experiment HW1 shows
   the best accuracy results. Also, some homeworks thrive well with just using training data from the previous homework (PreviousOneHW-based, e.g. HW3). 
 %
%




Figure \ref{fig:figure10} shows the comparison of 
prediction results (RMSE) 
for IntroCN of our method and baseline 
  with increasing number of regression models. For
  a single PLMR model, the 
  Meanscore baseline has better performance. But as the number of personalized models increases, the PLMR outperforms the baseline approach.

\subsubsection{Comparative Performance}

Table \ref{tab:table1}
shows the comparison between baseline  approach (Meanscore) and 
the predictive model
for the PreviousHW-based experiments for  the  AllStMed group. We 
  cannot report results for the KT-IDEM model 
  since, it solves  the binary classification problem only.
Table \ref{tab:table3} shows the comparison of the accuracy and F1 scores of the AllStLearn groups 
with baseline approaches.   We notice that for predicting the second homework, which only 
uses the information from HW1, the predictive model  is not as good as
the mean baseline, which reflects that under the 
situation of lack of 
necessary amount of information, linear regression models cannot 
always outperform the baseline. But as the  dataset gets larger, 
our approach outperforms the baseline due to the availability of more 
training data.  From Table \ref{tab:table3}, we also notice for some homework, KT-IDEM has better performance than PLMR 
(HW7 and HW4). This could be due to unstable academic activities during these two study 
periods, which can effect the performance of PLMR. However, for most of the situation, our model can gain better prediction performance.

Table \ref{tab:table5} shows the comparison of PreviousOneHW-based prediction results of AllStLearn group. With 
limited information, i.e. using only the  previous one homework's information, our PLMR 
approach outperforms the KT-IDEM baseline.

\subsubsection{Feature Importance}

 We  test the effect of each feature set 
in predicting the assessment scores by training the models under the absence 
of each feature group. 
%
%
For the StLearn 
course, since there is no limit on homework attempts, we do not add Interval-Based 
feature groups to the
predictive model. Figure \ref{fig:figure7} shows the comparison of each 
prediction result for AllStMed, PartialStMed, AllStLearn and PartialStLearn cohorts. 

Analyzing these results we observe that for the StLearn MOOC, the meanscore is a significant 
feature and removing it leads to a substantial 
decrease in the accuracy results for the All and Partial- cohorts. 
For the AllStMed MOOC the removal of session features leads to the most decrease in performance (i.e., increased 
RMSE). This suggests that features related to the sessions which capture student engagement are crucial for predicting 
the final homework scores. For the PartialStMed MOOC, the use of all feature types or a subset does not show a clear winner. This could 
be due the varying characteristics of students within these group. 

Another way to analyze feature importance is to exclude the influence of 
meanscore which is a dominant feature in predicting a 
student's future performance. The 
evaluation formula of the importance of 
the $i_{th}$ feature (excluding meanscore feature) is as follows:

\begin{equation}\label{eq6}
\begin{aligned}
\begin{split}
I_i=\frac{1}{N} \sum_{n=1}^{N} \frac{\sum_{d=1}^{l}|p_{n_S,d}f_{n_S,i}w_{d,i}|}{\sum_{d=1}^{l} | p_{n_S,d}\sum_{k=1}^{n_F}f_{n_S,k}w_{d,k}|},
\end{split}
\end{aligned}
\end{equation}

where $N$ is number of test samples, $n_S$ is the student number corresponding to the $n_{th}$ test sample.  $f_{n_S,i}$ is the feature value of an interaction between student $n_S$ and activity $i$. $n_F$ is the number of features. $l$ is the number of 
linear regression models. $w_{d,i}$ is the coefficient of $d_{th}$ linear 
regression model with $i_{th}$ feature, and $p_{n_S,d}$ is the 
membership of student $n_S$ with the $d_{th}$ regression model.
We calculate each feature's 
importance by calculating the percentage contribution 
of each 
feature to the overall grade prediction.

\begin{figure}[th]
\centering
  \includegraphics[width=0.7\columnwidth]{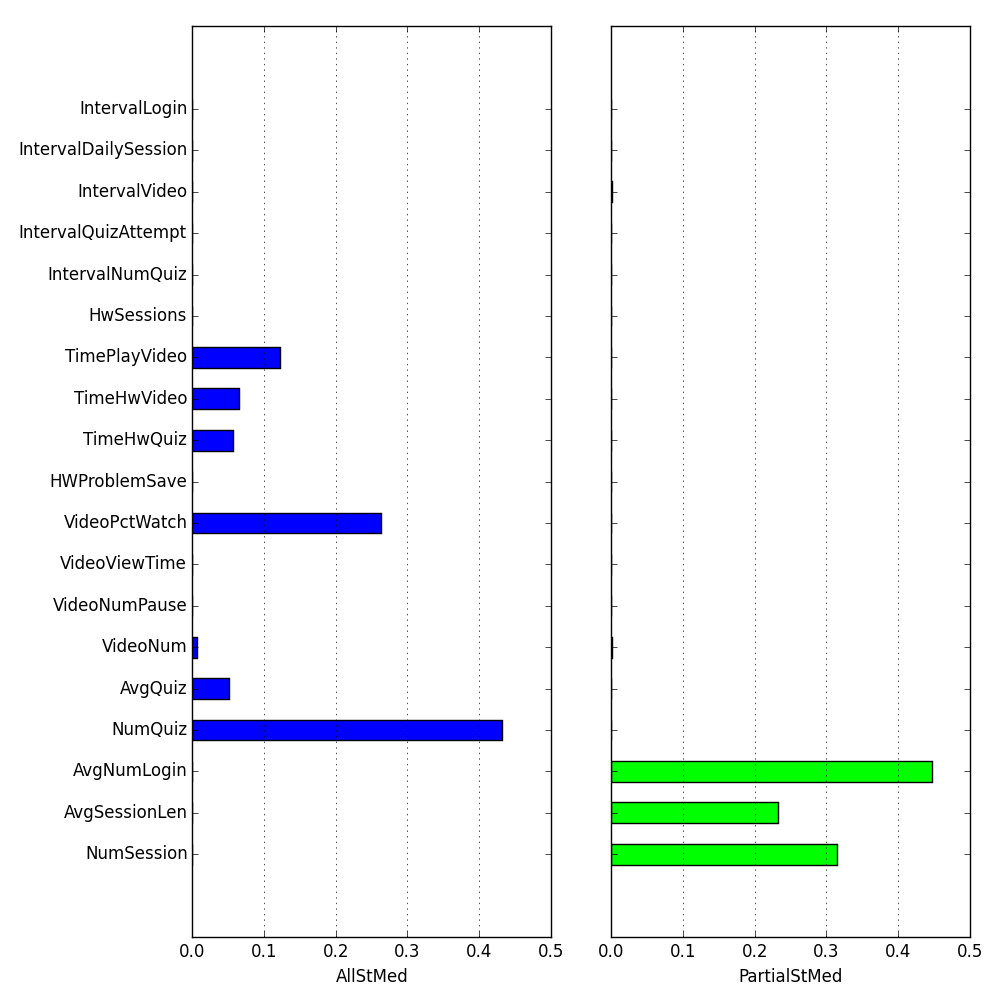}
  \caption{Feature importance for StMed.~\label{fig:figure9}}
\vspace{-1em}
\end{figure}

Figure \ref{fig:figure9} shows the feature importance on AllStMed and PartialStMed group, excluding Meanscore feature. We can see these two groups have completely different feature importance. \textbf{NumQuiz} and \textbf{VideoPctWatch} are the most important for AllStMed group besides \textbf{Meanscore} feature while all the Session features are important for PartialStMed group.





\section{CONCLUSION AND FUTURE WORK}

In this work we formulated a personalized multiple linear regression model 
to predict the homework/exam grades for a student enrolled and participating within a MOOC. Our contributions 
include engineering features 
that capture a student's studying behavior and learning habits, derived solely 
from the server logs of MOOCs. 

We evaluated our framework on three OpenEdX MOOC courses provided by an initiative at Stanford University. Our 
experimental evaluation shows improved performance in terms of prediction of real time homework scores when 
compare to baseline methods. We also studied on different groups of student participants according to their motivation and representation, some who 
complete all the assessments and some who only finish a subset of the provided assignments. Features associated 
with engagement (logging multiple times), studying materials (viewing videos and attempting quizzes) were found 
to be important along with prior homework scores for this prediction problem. 

Given, 
the large number of users it is extremely hard to
monitor the progress of users and provide them 
with individualized feedback. If MOOCs are to move 
beyond being a content repository, the ability to 
guide users through the course successfully is essential. For 
this we need to know when to intervene and how to 
be productive in our intervention. 
In the future, we seek to use this formulation within a real-time early warning or intervention system that will seek 
to improve student retention and improve their overall performance.

%




\bibliography{mybib}{}
\bibliographystyle{plain}








\balancecolumns
\end{document}